\documentclass[%
preprint,%
secnumarabic,%
amssymb, amsmath,aip,
apsrev4-1,%apl,%
superscriptaddress,%
frontmatterverbose,
showpacs,
showkeys
]{revtex4-1}

\usepackage{multirow}
\usepackage{array}
\usepackage{amssymb}
\usepackage{amsthm}
\usepackage{amsmath}
\usepackage{xfrac}
\usepackage{amsfonts}
\usepackage{graphicx}
\usepackage{steinmetz} % angle symbol using phasor notation
\usepackage{hyperref}
\usepackage[dvipsnames]{xcolor}
\usepackage{bm} %boldsymbol
\usepackage{ulem,xpatch} % to strikethrough
\usepackage[percent]{overpic}
\usepackage{tikz}
\usepackage{enumitem}
\usepackage{xfrac}
\usepackage{mathtools}

\definecolor{Blue}{rgb}{0,0,0.898}
\definecolor{Green}{rgb}{0,0.506,0}
\definecolor{Green2}{rgb}{0,0.584,0}
\definecolor{Green3}{rgb}{0,0.502,0}
\definecolor{Red}{rgb}{0.898,0,0}
\definecolor{Magenta}{rgb}{0.816,0,0.816}
\definecolor{Gray1}{rgb}{0.502,0.502,0.502}
\definecolor{Gray1}{rgb}{0.388,0.388,0.388}
\definecolor{Gray2}{rgb}{0.702,0.702,0.702}
\definecolor{Gray3}{rgb}{0.6235,0.6235,0.6235}

\newcommand*{\bdash}[1][3em]{\rule[0.5ex]{#1}{1.25pt}}
\newcommand{\sgn}{\operatorname{sgn}} 

\def\emcirc{\setbox0\hbox{\bdash[1.5em]}%
	\rlap{\hbox to \wd0{\hss$\circ$\hss}}\box0}
\def\emast{\setbox0\hbox{\bdash[1.5em]}%
	\rlap{\hbox to \wd0{\hss$\boldsymbol{\ast}$\hss}}\box0}
\def\emdot{\setbox0\hbox{\bdash[1.5em]}%
	\rlap{\hbox to \wd0{\hss\textbullet\hss}}\box0}
\def\emdiam{\setbox0\hbox{\bdash[1.5em]}%
	\rlap{\hbox to \wd0{\hss$\boldsymbol{\diamond}$\hss}}\box0}

\begin{document}
	%\begin{CJK*}{UTF8}{} % Use default fonts from CJK
	\title{Application of the method of multiple scales to unravel energy exchange in nonlinear locally resonant metamaterials}%
	
	\author{P B Silva}
	\email{p.brandao.silva@tue.nl}
	\affiliation{Eindhoven University of Technology\\
		PO Box 513, 5600 MB\\ Eindhoven, The Netherlands }%
	\author{M J Leamy}
	\affiliation{
		George W. Woodruff School of Mechanical Engineering,\\ Georgia Institute of Technology,\\
		Atlanta, GA 30332-0405
	}
	\author{M G D Geers}
	\affiliation{Eindhoven University of Technology\\
		PO Box 513, 5600 MB\\ Eindhoven, The Netherlands }%
	\author{V G Kouznetsova}
	\affiliation{Eindhoven University of Technology\\
		PO Box 513, 5600 MB\\ Eindhoven, The Netherlands }%
	%\date{\today}
	
	\begin{abstract}
		In this paper, the effect of weak nonlinearities in 1D locally resonant metamaterials is investigated via the method of multiple scales. Commonly employed to the investigate the effect of weakly nonlinear interactions on the free wave propagation through a phononic structure or on the dynamic response of a Duffing oscillator, the method of multiple scales is here used to investigate the forced wave propagation through locally resonant metamaterials. The perturbation approach reveals that energy exchange may occur between propagative and evanescent waves induced by quadratic nonlinear local interaction.   
	\end{abstract}
	\keywords{Nonlinear Wave Dynamics; Energy Exchange; Local Resonance; Wave Propagation} 
	\maketitle
\section{Introduction}
Locally resonant metamaterials and phononic crystals have been receiving considerable attention due to the ``exotic'' dynamic features they may exhibit by engineering a unit cell of the periodic structure. Induced by distinct physics, these structures exhibit band gaps, {\it i.e.} frequency zones in which certain waves cannot propagate, which make them good sound and vibration insulators. In addition, the associated translational symmetry and effective dynamic properties of these periodic structures have been shown to provide unique possibilities for wave manipulation, providing opportunities to break through the traditional limits in imaging, wave guiding and focusing \cite{Hussein2014}. Recently, the consideration of nonlinearity in such structures has been shown to induce amplitude-dependent spectral characteristics \cite{Manktelow2014, Narisetti2011} as well as break time-reversal symmetry \cite{Fleury2014}. 

Most of the aforementioned works have considered nonlinear neighbouring interactions rather than nonlinear local interactions. Within the nonlinear dynamics community, a single essentially nonlinear attachment to a structure, the so-called nonlinear energy sink (NES), has been shown to induce irreversible energy transfer mechanisms \cite{Vakakis2009}. The effect of a periodic distribution of nonlinear resonators has been less investigated. Works in this direction have shown the amplitude-dependent dispersion behavior of these structures and the effect of chaotic regimes on the bandwidth of the attenuation zone characteristic of locally resonant metamaterials \cite{Fang2017, Lazarov2007}. 

In this work, we investigate discrete metamaterials with nonlinear local interactions. Asymmetric nonlinear interactions of quadratic and quadratic-cubic types are considered and used as approximation of the material nonlinearities in real applications. The method of multiple scales is used to model the wave-wave interaction induced by the nonlinearity in the mechanical systems. Their responses under external excitation for several amplitude levels are analyzed and validated with direct numerical simulations. 

\section{Mechanical model system}
\begin{figure}[h!]
	\centering
	\includegraphics[width=\textwidth]{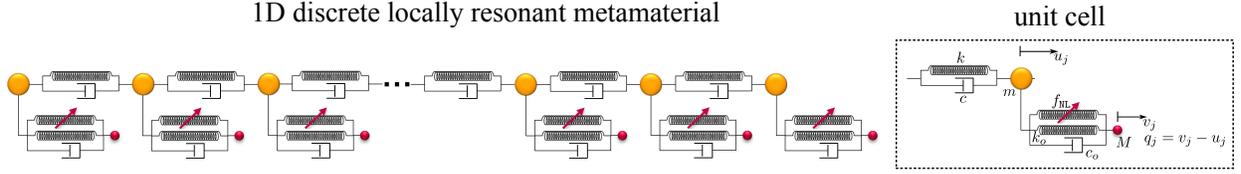}
	\caption{Mechanical model system.}
	\label{fig:discrete_f1}
\end{figure}

A one-dimensional discrete locally resonant metamaterial lattice as shown in Fig. \ref{fig:discrete_f1} is used to investigate the energy exchange between wave modes when the local interaction is nonlinear. Inspired by fact that any nonlinear function can be expanded in Taylor series as, for instance, nonlinear functions describing realistic nonlinear material models, and that low-order approximations are able to describe complex energy exchange phenomena due to realistic material nonlinearity \cite{Silva2018}, the analysis in this paper is limited to nonlinear interactions of quadratic type.    

The equations of motion for a unit cell of the metamaterial model under no external applied loads are given by:
\begin{subequations}
	\begin{align}
	&\begin{array}{ll}
	m\ddot{u}_j+c\left(-\dot{u}_{j-1}+2\dot{u}_j-\dot{u}_{j+1}\right)+c_o(\dot{u}_j-\dot{v}_j)+k\left(-{u}_{j-1}+2{u}_j-{u}_{j+1}\right)+\cdots&~\\
	+k_o(u_j-v_j)+\gamma_2 k_o(u_j-v_j)^2=0 ,
	\end{array}\\
	&
	\begin{array}{ll}
	m_o\ddot{v}_j+c_o(\dot{v}_j-\dot{u}_j)+k_o(v_j-u_j)-\gamma_2 k_o(v_j-u_j)^2=0,
	\end{array}
	\end{align}
	\label{eq:discrete_e1}
\end{subequations}
where $u_j$ and $v_j$ are the displacements of the main chain unit and local oscillator, respectively, $(\dot{~})$ is used to denote the differentiation with respect to time $t$, $m$ and $m_o$ are, respectively, the masses of the main chain unit and local oscillator, $k$ is the spring stiffness connecting the chain masses, $c_o$ is the damping constant of the dash-pot between the local oscillator and the chain mass, $k_o$ is the stiffness of the spring connecting the local oscillator to the main chain mass, $\gamma_2$ is the nonlinear coefficient relative to the quadratic term, respectively.

In terms of the non-dimensional parameters, {\it i.e.} normalized displacements $\bar{u}_j = \sfrac{{u}_j}{L_0}$ and $\bar{v}_j = \sfrac{{v}_j}{L_0}$, with $L_0$ being the natural length of the spring connecting the local oscillator to the main chain unit, the angular frequencies $\omega_m = \sfrac{k}{m}$ and $\omega_{\tt R} = \sfrac{k_o}{m_o}$, damping factors $\zeta_m = \sfrac{c}{2\omega_m m}$ and $\zeta_o = \sfrac{c_o}{2\omega_{\tt R} m_o}$, associated to the main chain unit mass and the local oscillator, respectively, and mass fraction $\beta = \sfrac{m_o}{m}$, the governing equations are rewritten as:
\begin{subequations}
	\begin{align}
	&\begin{array}{ll}
	\ddot{\bar{u}}_j+2\zeta_m\omega_m\left(-\dot{\bar{u}}_{j-1}+2\dot{\bar{u}}_j-\dot{\bar{u}}_{j+1}\right)+\omega_m^2\left(-{\bar{u}}_{j-1}+2{\bar{u}}_j-{\bar{u}}_{j+1}\right)+\cdots&~\\
	+2\beta\zeta_o\omega_{\tt R}(\dot{\bar{u}}_j-\dot{\bar{v}}_j)+\beta \omega_{\tt R}^2({\bar{u}}_j-{\bar{v}}_j)+\gamma_2 \beta \omega_{\tt R}^2({\bar{u}}_j-{\bar{v}}_j)^2=0 ,
	\end{array}\\
	&
	\begin{array}{ll}
	\ddot{\bar{v}}_j+2\zeta_o\omega_{\tt R}(\dot{\bar{v}}_j-\dot{\bar{u}}_j)+\omega_{\tt R}^2({\bar{v}}_j-{\bar{u}}_j)-\gamma_2 \omega_{\tt R}^2({\bar{v}}_j-{\bar{u}}_j)^2=0.
	\end{array}
	\end{align}
	\label{eq:discrete_e2}
\end{subequations}
Then, rewriting the equations of motion in terms of the dimensionless continuous time variable $\tau = \omega_m t$, it yields:
\begin{subequations}
	\begin{align}
	&\begin{array}{ll}
	\ddot{\bar{u}}_j+2\zeta_m\left(-\dot{\bar{u}}_{j-1}+2\dot{\bar{u}}_j-\dot{\bar{u}}_{j+1}\right)+\left(-{\bar{u}}_{j-1}+2{\bar{u}}_j-{\bar{u}}_{j+1}\right)+\cdots&~\\
	+2\beta\zeta_o\bar{\omega}_{\tt R}(\dot{\bar{u}}_j-\dot{\bar{v}}_j)+\beta \bar{\omega}_{\tt R}^2({\bar{u}}_j-{\bar{v}}_j)+\gamma_2 \beta \bar{\omega}_{\tt R}^2({\bar{u}}_j-{\bar{v}}_j)^2=0 ,
	\end{array}\\
	&
	\begin{array}{ll}
	\ddot{\bar{v}}_j+2\zeta_o\bar{\omega}_{\tt R}(\dot{\bar{v}}_j-\dot{\bar{u}}_j)+\bar{\omega}_{\tt R}^2({\bar{v}}_j-{\bar{u}}_j)-\gamma_2 \bar{\omega}_{\tt R}^2({\bar{v}}_j-{\bar{u}}_j)^2=0,
	\end{array}
	\end{align}
	\label{eq:discrete_e3}
\end{subequations}
where $\bar{\omega}_{\tt R} = \sfrac{\omega_{\tt R}}{\omega_m}$ is a ratio of angular frequencies. 

Considering a dimensionless continuous time variable $\tau = \sqrt{\beta}{\omega}_{\tt R}t$, the equations of motion become:
\begin{subequations}
	\begin{align}
	&\begin{array}{ll}
	\ddot{\bar{u}}_j+2\zeta_m\gamma_c\left(-\dot{\bar{u}}_{j-1}+2\dot{\bar{u}}_j-\dot{\bar{u}}_{j+1}\right)+\gamma_c^2\left(-{\bar{u}}_{j-1}+2{\bar{u}}_j-{\bar{u}}_{j+1}\right)+\cdots&~\\
	+2\zeta_o\sqrt{\beta}(\dot{\bar{u}}_j-\dot{\bar{v}}_j)+({\bar{u}}_j-{\bar{v}}_j)+\gamma_2({\bar{u}}_j-{\bar{v}}_j)^2=0 ,
	\end{array}\\
	&
	\begin{array}{ll}
	\beta\ddot{\bar{v}}_j+2\zeta_o\sqrt{\beta}(\dot{\bar{v}}_j-\dot{\bar{u}}_j)+({\bar{v}}_j-{\bar{u}}_j)-\gamma_2({\bar{v}}_j-{\bar{u}}_j)^2=0,
	\end{array}
	\end{align}
	\label{eq:discrete_e3}
\end{subequations}
where $\gamma_c = \omega_m{\left(\sqrt{\beta} \bar{\omega}_{\tt R}\right)}^{-1}$ is a ratio of angular frequencies. 

\subsection{Associated conservative linear dynamical system}
The associated conservative linear dynamical system is given by:
\begin{subequations}
	\begin{align}
	&\begin{array}{ll}
	\ddot{\bar{u}}_j+\gamma_c^2\left(-\bar{u}_{j-1}+2\bar{u}_j-\bar{u}_{j+1}\right)
	+(\bar{u}_j-\bar{v}_j)=0 ,
	\end{array}\\
	&
	\begin{array}{ll}
	\beta\ddot{\bar{v}}_j+(\bar{v}_j-\bar{u}_j)=0,
	\end{array}
	\end{align}
	\label{eq:discrete_linear_e1}
\end{subequations}
which, in matrix form, writes:
\begin{equation}
	{\bf M}\ddot{\bar{\bf u}}_j+\gamma_c^{2}{\displaystyle \sum_{p = -1}^{1}}{\bf K}_p {\bar{\bf u}}_j = {\bf 0},
	\label{eq:discrete_linear_e1b}
\end{equation}
with 
\begin{equation}
	\begin{array}{llllllll}
	{\bf M} &= 
	\begin{bmatrix}
		1 & 0\\
		0 & \beta
	\end{bmatrix}, & {\bf K}_0 &= 
	\begin{bmatrix}
		2+\gamma_c^{-2} & -\gamma_c^{-2}\\
		-\gamma_c^{-2} & \gamma_c^{-2}
	\end{bmatrix},&
	{\bf K}_{-1} = {\bf K}_1&= 
	\begin{bmatrix}
		-1 & 0\\
		0 & 0
	\end{bmatrix}, & \bar{\bf u}_j &= 
	\begin{bmatrix}
		\bar{u}_j\\
		\bar{v}_j
	\end{bmatrix}.
	\end{array}	
\end{equation}
The solution of this system might be given by a time-harmonic wave solution of the form $\bar{\bf u}_j(\tau) = \tilde{\bf u}_j(\bar{\omega})+\tilde{\bf u}_j(-\bar{\omega})$, where $\bar{\omega} = \sfrac{\omega}{\sqrt{\beta}\omega_{\tt R}}$ is the normalized angular frequency,
\begin{equation}
	\tilde{\bf u}_j(\bar{\omega})={\bf U}_j(\bar{\omega})e^{{\tt i}(\bar{\omega} \tau -\mu j)}
	\label{eq:discrete_linear_e2a}
\end{equation}
and
\begin{equation}
	\tilde{\bf u}_j(-\bar{\omega})=\tilde{\bf u}_j^{\ast}(\bar{\omega}),
	\label{eq:discrete_linear_e2b}
\end{equation}
to comply with the time reverse symmetry property \cite{Joannopoulos2008}; the superscript $\ast$ standing for the complex conjugate. Due to the periodic feature of the linear metamaterial, Bloch-Floquet conditions might be satisfied, {\it i.e.}:
\begin{equation}
	\tilde{\bf u}_{j+1} = \tilde{\bf u}_j e^{-{\tt i} \mu}.		 				\label{eq:discrete_linear_e3a}
\end{equation}
Herein, $\mu$ is the dimensionless wavenumber, $\tilde{\bf u}_j$ is the vector of displacements $[\tilde{u}_j, \: \tilde{v}_j]^T$ relative to the unit cell $j$. For the mechanical model under concern, with a single degree-of-freedom in the main chain per unit cell, this implies:
\begin{equation}
	{\bf U}_{j+1}(\bar{\omega}) = {\bf U}_j(\bar{\omega}),
	\label{eq:discrete_linear_e3b}
\end{equation}
where ${\bf U}_j$ is the vector of wave mode amplitudes $[U_j, \: V_j]^T$ relative to the unit cell $j$, and the wave mode shapes are constant  functions of the normalized angular frequency. As a result, the subscript $j$ can be dropped from the wave mode shapes in what follows.

Substituting the time-harmonic solution \eqref{eq:discrete_linear_e2a} in Eq. \eqref{eq:discrete_linear_e1b}, it yields:
\begin{equation}
{\bf D}_{\tt L}(\bar{\omega},\mu){\bf U} = {\bf 0}, 
\label{eq:discrete_linear_e5}
\end{equation}
where
\begin{equation}
{\bf D}_{\tt L}(\bar{\omega},\mu) = 
\begin{bmatrix}
\left(-\bar{\omega}^2 + 2\gamma_c^2\left(1-\cos{\mu}\right)+1\right) & -1\\
-1 & \left(-\beta\bar{\omega}^2 + 1\right)
\end{bmatrix}.
\label{eq:discrete_linear_e6}
\end{equation}
Non-trivial solutions of Eq. \eqref{eq:discrete_linear_e5} are obtained by forcing the determinant of ${\bf D}_{\tt L}(\bar{\omega},\mu)$ to vanish, which yields the following dispersion relation:
\begin{equation}
\cos \mu = 1 - \frac{\bar{\omega}^2}{2\gamma_c^2}{\frac{\beta\bar{\omega}^2-(\beta+1)}{\beta\bar{\omega}^2-1}},
\label{eq:discrete_linear_e7}
\end{equation}
and the following relation between the wave mode amplitudes $U$ and $V$:
\begin{equation}
U = \left(1 - \beta\bar{\omega}^2\right) V.
\label{eq:discrete_linear_e8}
\end{equation}
Then, the normalized wave mode shape vector $\boldsymbol{\Phi}$ is given by:
\begin{equation}
\boldsymbol{\Phi}(\bar{\omega}) = \frac{1}{\sqrt{1+\left(1-\beta\bar{\omega}^2\right)^2}}
\begin{bmatrix}
1-\beta\bar{\omega}^2\\
1
\end{bmatrix}.
\label{eq:discrete_linear_e9}
\end{equation}
Alternatively, one can use \eqref{eq:discrete_linear_e8} to rewrite \eqref{eq:discrete_linear_e5} in scalar form, as follows:
\begin{equation}
{D}_{\tt L}(\bar{\omega},\mu) U= 0,	
\label{eq:discrete_linear_e10}
\end{equation}
with
\begin{equation}
{D}_{\tt L}(\bar{\omega},\mu) = -\bar{\omega}^2 + 2\gamma_c^2\left(1-\cos{\mu}\right)+\frac{1}{\beta\bar{\omega}^2-1}.
\label{eq:discrete_linear_e11}
\end{equation}
Non-trivial solutions of Eq. \eqref{eq:discrete_linear_e10} are obtained by forcing $D_{\tt L}(\bar{\omega},\mu)$ to vanish, which recovers the dispersion relation given by Eq. \eqref{eq:discrete_linear_e7}.

\section{Analytical modeling via the method of multiple scales}
Herein, the method of multiple scales is used to obtain an approximative solution for the weakly nonlinear locally resonant metamaterial under concern in the cases of free and forced wave propagation. The main objective is to get a physical and mathematical description of the wave-wave interaction in locally resonant metamaterials with nonlinear local oscillators. 

In general, the method of multiple scales (MMS) seeks for a uniformly valid solution in the form of an asymptotic expansion. In this section, a solution by the method of multiple scales is sought for the wave propagation problem:
\begin{subequations}
	\begin{align}
	&\begin{array}{ll}
	\ddot{\bar{u}}_j+2\zeta_m\gamma_c\left(-\dot{\bar{u}}_{j-1}+2\dot{\bar{u}}_j-\dot{\bar{u}}_{j+1}\right)+\gamma_c^2\left(-{\bar{u}}_{j-1}+2{\bar{u}}_j-{\bar{u}}_{j+1}\right)+\cdots&~\\
	+2\zeta_o\sqrt{\beta}(\dot{\bar{u}}_j-\dot{\bar{v}}_j)+({\bar{u}}_j-{\bar{v}}_j)+\gamma_2({\bar{u}}_j-{\bar{v}}_j)^2=0 ,
	\end{array}\\
	&
	\begin{array}{ll}
	\beta \ddot{\bar{v}}_j+2\zeta_o\sqrt{\beta}(\dot{\bar{v}}_j-\dot{\bar{u}}_j)+({\bar{v}}_j-{\bar{u}}_j)-\gamma_2({\bar{v}}_j-{\bar{u}}_j)^2=0.
	\end{array} 
	\end{align}
	\label{eq:MMS_e1}
\end{subequations}
In matrix form, the equations of motion can be written as:
\begin{equation}
	{\bf M}\ddot{\bar{\bf u}}_j+ 2\zeta_m\gamma_c {\displaystyle\sum_{p = -1}^{1}}{\bf C}_p \dot{\bar{\bf u}}_{j+p}+ \gamma_c^2 {\displaystyle \sum_{p = -1}^{1}}{\bf K}_p {\bar{\bf u}}_{j+p} + \gamma_2 \boldsymbol{\Phi}_{f} \left(\bar{u}_j-\bar{v}_j\right)^2= {\bf 0},
	\label{eq:MMS_e1b}
\end{equation}
with 
\begin{equation}
	{\bf C}_0 = \begin{bmatrix}
		2+\bar{\zeta}\omega^{-1}_{\tt R} & -\bar{\zeta}\omega^{-1}_{\tt R}\\
		-\bar{\zeta}\omega^{-1}_{\tt R} & \bar{\zeta}\omega^{-1}_{\tt R}
	\end{bmatrix},
	{\bf C}_{-1} = {\bf C}_{1} = \begin{bmatrix}
		-1 & 0\\
		0 & 0
	\end{bmatrix},
	\boldsymbol{\Phi}_{f} = \begin{bmatrix}
	1\\
	-1
	\end{bmatrix}.
\end{equation}
with $\bar{\zeta} = \sfrac{\zeta_o}{\zeta_m}$.

The interest in this paper is in describing wave-wave interaction between a propagative and an evanescent wave, which has localized spatial effect. Therefore, the perturbation solution might involve not only multiple time scales, but also multiple space scales might be considered. 

Let $\varepsilon \ll 1$ be small bookkeeping parameters. First-order expansion is considered in both time and space scales, and these scales are defined as:
\begin{subequations}
	\begin{align}	
		T_0 = \tau, \quad T_1 = \varepsilon \tau,\\
		X_0 = j, \quad X_1 = \varepsilon j. 
	\end{align}
	\label{eq:MMS_e2}
\end{subequations}
Since $\varepsilon \ll 1$, each time (resp. space) scale is slower (resp. longer) than its predecessor. 

Within the framework of the method of multiple scales, the displacement solution for an arbitrary unit cell $j$ ($\bar{\bf u}_j$) can be written as an asymptotic expansion of the form:
\begin{equation}
\bar{\bf u}_j \simeq \varepsilon \hat{\bf u}_j^{(0)}+\varepsilon^2 \hat{\bf u}_j^{(1)}+ \cdots,
\label{eq:MMS_e5}
\end{equation}
where $\hat{\bf u}_j^{(n)} = \hat{\bf u}_j^{(n)}(X_0,X_1,T_0,T_1)$ is the normalized displacement at the order $n$. 

As a consequence of the consideration of multiple time scales, the time derivatives also involve several orders of perturbation, as follows:
\begin{subequations}
	\begin{align}	
		(\dot{~}) =& D_{T_0} + \varepsilon D_{T_1}  + \varepsilon^2 D_{T_2} + \cdots,\\
		(\ddot{~}) =& D^2_{T_0} + \varepsilon 2D_{T_0}D_{T_1}  + \varepsilon^2 \left(D^2_{T_2}+2D_{T_2}D_{T_0}\right) + \cdots,
	\end{align}
	\label{eq:MMS_e3}
\end{subequations}
where $D^m_{T_n} = \sfrac{\partial^m ~}{\partial T_n^m}$. Analogously, for the discrete model system, the relation between neighbouring degrees-of-freedom might also be expressed in terms of several orders of space scales, as follows:
\begin{subequations}
	\begin{align}	
		{u}_{j+1}^{(n)}-{u}_{j}^{(p)} =& \tilde{u}^{(n)}(T_0,T_1) \left\{D_{X_0}\left(e^{-{\tt i}\mu_0 X_0}\right) + \varepsilon D_{X_1}\left(e^{-{\tt i}\mu_1 X_1}\right)\right\} + \cdots,\\
		{u}_{j+1}^{(n)}-2{u}_{j}^{(p)}+{u}_{j-1}^{(p)} =& \tilde{u}^{(n)}(T_0,T_1) \left\{D^2_{X_0}\left(e^{-{\tt i}\mu_0 X_0}\right) + \varepsilon 2D_{X_0}\left(e^{-{\tt i}\mu_0 X_0}\right)D_{X_1}\left(e^{-{\tt i}\mu_1 X_1}\right)\right\} + \cdots,
	\end{align}
	\label{eq:MMS_e4}
\end{subequations}
with $n = \left\{0, 1\right\}$. A detailed derivation of these expressions is provided in Appendix \ref{app:A}.

In what follows, the damping factor $\bar{\zeta}$ and the quadratic nonlinear coefficient $\gamma_2$ are scaled as:
\begin{equation}
	\zeta_m = \varepsilon \hat{\zeta}_m, \quad {\gamma}_2 = \varepsilon^0 \hat{\gamma}_2,
	\label{eq:MMS_e6}	
\end{equation}
such that the damping force appear at the same order of the quadratic nonlinear interaction. Then, substituting Eqs. \eqref{eq:MMS_e2}-\eqref{eq:MMS_e6} into Eq. \eqref{eq:MMS_e1b}, and collecting matching orders of $\varepsilon$, yields:
%\begin{subequations}
%	\begin{align}
%	\varepsilon^1:\quad&
%	{\bf M} D_{T_0}^2 \hat{\bf u}^{(0)}_{j}+ \gamma_c^2 {\displaystyle \sum_{p = -1}^{1}}{\bf K}_p {\hat{\bf u}}^{(0)}_{j+p}= {\bf 0}, &
%	\label{eq:MMS_e7a}\\
%	\varepsilon^2:\quad&
%    {\bf M}D_{T_0}^2 \hat{\bf u}^{(1)}_{j}+ \gamma_c^2 {\displaystyle \sum_{p = -1}^{1}}{\bf K}_p {\hat{\bf u}}^{(1)}_{j+p} = -2{\bf M}D_{T_1}D_{T_0} \hat{\bf u}^{(0)}_{j}-2\hat{\zeta}_m\gamma_c {\displaystyle\sum_{p = -1}^{1}}{\bf C}_p D_{T_0}{\hat{\bf u}}^{(0)}_{j+p}&\\ \nonumber
%    &-2\gamma_c^2D_{X_0}\left(e^{-{\tt i}\mu_0 X_0}\right)\underbrace{\displaystyle \sum_{p = -1}^{0}{\bf K}_p^{1}{\tilde{\bf u}}^{(0)}_{j+p}}_{D_{X_1}\left(e^{-{\tt i}\mu_1 X_1}\right)\tilde{u}(T_0,T_1)}- \hat{\gamma}_2 \boldsymbol{\Phi}_{f} \left(\hat{u}^{(0)}_{j}-\hat{v}^{(0)}_{j}\right)^2.
%    \label{eq:MMS_e7b}
%	\end{align}	
%	\label{eq:MMS_e7}
%\end{subequations}
\begin{subequations}
	\begin{align}
	\varepsilon^1:\quad&
	{\bf M} D_{T_0}^2 \hat{\bf u}^{(0)}_{j}+ \gamma_c^2 {\displaystyle \sum_{p = -1}^{1}}{\bf K}_p {\hat{\bf u}}^{(0)}_{j+p}= {\bf 0}, &
	\label{eq:MMS_e7a}\\
	\varepsilon^2:\quad&
    {\bf M}D_{T_0}^2 \hat{\bf u}^{(1)}_{j}+ \gamma_c^2 {\displaystyle \sum_{p = -1}^{1}}{\bf K}_p {\hat{\bf u}}^{(1)}_{j+p} = -2{\bf M}D_{T_1}D_{T_0} \hat{\bf u}^{(0)}_{j}-2\hat{\zeta}_m\gamma_c {\displaystyle\sum_{p = -1}^{1}}{\bf C}_p D_{T_0}{\hat{\bf u}}^{(0)}_{j+p}\label{eq:MMS_e7b}&\\ \nonumber
    &-2\gamma_c^2D_{X_0}\left(e^{-{\tt i}\mu_0 X_0}\right){D_{X_1}\left(e^{-{\tt i}\mu_1 X_1}\right)\tilde{\bf u}^{(0)}_{j}}- \hat{\gamma}_2 \boldsymbol{\Phi}_{f} \left(\hat{u}^{(0)}_{j}-\hat{v}^{(0)}_{j}\right)^2.
	\end{align}	
	\label{eq:MMS_e7}
\end{subequations}\linebreak
Herein,
\begin{equation}
	{\bf K}_0^1 = 
	\begin{bmatrix}
	1 & 0\\
	0 & 0
	\end{bmatrix}, \quad
	{\bf K}_{-1}^1 = 
	\begin{bmatrix}
	-1 & 0\\
	0 & 0
	\end{bmatrix}.	 
\end{equation}
In order to investigate the importance of the wave-wave interaction between the primary and subharmonic wave modes, it is convenient to assume that the zeroth-order solution $\hat{\bf u}^{(0)}_j$ in Eq. \eqref{eq:MMS_forced_e7a} is given by a superposition of these two wave modes, as follows:
\begin{equation}
\begin{array}{ll}
\hat{\bf u}^{(0)}_j =& \underbrace{\hat{U}^{(p)}(X_1,T_1)  \boldsymbol{\Phi}_p e^{{\tt i}(\bar{\omega}_p T_0 - \mu_p X_0)}}_{\hat{\bf u}^{(0,p)}_j}+\underbrace{\hat{U}^{(s)}(X_1,T_1) \boldsymbol{\Phi}_s e^{{\tt i}(\bar{\omega}_s T_0 - \mu_s X_0)}}_{\hat{\bf u}^{(0,s)}_j}+\text{c.c.}
\end{array}
\label{eq:app_u_0order}
\end{equation}
where ${\tt i}$ is the imaginary unit, $\boldsymbol{\Phi}_p$ and $\boldsymbol{\Phi}_s$ are the vectors of wave mode shapes, and $\hat{U}^{(p)}(X_1,T_1)$, $\hat{U}^{(s)}(X_1,T_1)$ are the complex wave amplitude functions of the slow and long time and space scales corresponding to the linear waves $(\bar{\omega}_p,\mu_p)$ and $(\bar{\omega}_s,\mu_s)$, respectively. 
Herein, we assume the propagation of a primary wave mode $(\bar{\omega}_p, \mu_p)$ located in the optical branch of the corresponding linear metamaterial, with its angular frequency satisfying a $2:1$ relation with respect to the locally resonance frequency, {\it i.e.}
\begin{equation}
	\bar{\omega}_p = 2\bar{\omega}_{\tt R} + \varepsilon{\sigma}_p,
\label{eq:detuning_p}
\end{equation}
where $\sigma_p$ is a frequency detuning parameter. 

\begin{figure}[h!]
	\centering
	\includegraphics[width=.6\textwidth]{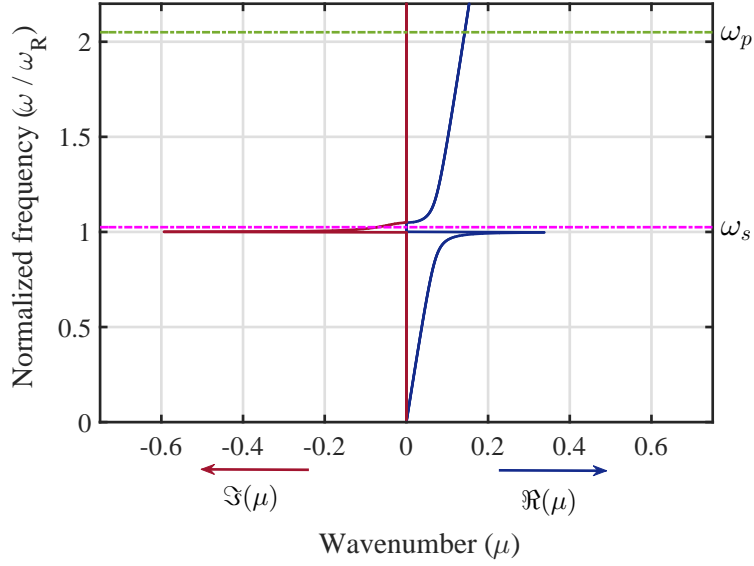}		
	\caption{Dispersion relation for the linear undamped locally resonant metamaterial depicting the wave modes selected for the wave-wave interaction analysis.} 
	\label{fig:dispersion_ps}
\end{figure}

Due to the nonlinearity, subharmonics and superharmonics of the primary wave mode can be generated. In particular, a half subharmonic wave mode $(\bar{\omega}_s, \mu_s)$ of evanescent nature can be induced since its associated frequency might approach the local resonant frequency by
\begin{equation}
	\bar{\omega}_s = \bar{\omega}_{\tt R} + \varepsilon{\sigma}_s,
\label{eq:detuning_s}
\end{equation}   
with $\sigma_s >0$ being a frequency detuning parameter. For the purpose of illustration, the dispersion diagram of the corresponding linear metamaterial depicting the primary propagating wave mode and an associated evanescent subharmonic wave mode is shown in Fig. \ref{fig:dispersion_ps}.  The primary and subharmonic wave modes to be considered within the multiple scales framework correspond to those of the approximate linear metamaterial. Therefore, the relation between the corresponding frequency and wavenumber is given by Eq. \eqref{eq:discrete_linear_e7}. From Eqs. \eqref{eq:detuning_p}, \eqref{eq:detuning_s} and \eqref{eq:discrete_linear_e7}, a closed-form relation for the analogous wavenumber detuning parameters can be expressed in terms of the angular frequencies and corresponding dispersion relations for the approximate linear metamaterial.

Substituting Eq. \eqref{eq:app_u_0order} into \eqref{eq:MMS_e7b}, yields:
\begin{equation} 
\begin{array}{l}
\displaystyle \sum_{m} \left\{{\bf M}D_{T_0}^2 \hat{\bf u}^{(1,m)}_{j}+ \gamma_c^2 {\displaystyle \sum_{p = -1}^{1}}{\bf K}_p {\hat{\bf u}}^{(1,m)}_{j+p}\right\} =  -2{\tt i} \displaystyle \sum_{m_0} \left\{\bar{\omega}_{m_0}{\bf M}D_{T_1} \hat{\bf u}^{(0,m_0)}_{j}-\mu_{m_0}\gamma_c^2\displaystyle \sum_{q = -1}^{0}{\bf K}_p^{1} {\hat{\bf u}}^{(0,m_0)}_{j+p}\right\}\\ 
    -2{\tt i}\hat{\zeta}_m\gamma_c\displaystyle \sum_{m_0}\bar{\omega}_{m_0} {\displaystyle\sum_{p = -1}^{1}}{\bf C}_p {\hat{\bf u}}^{(0,m_0)}_{j+p}- \hat{\gamma}_2 \boldsymbol{\Phi}_{f} \sum_{m_1} \hat{f}_2^{(m_1)}\left(\hat{q}^{(0,s)}_{j},\hat{q}^{(0,p)}_{j}\right),
\end{array}
	\label{eq:app_MMS_o2}
\end{equation}
where $m_0 = \left\{p,s\right\}$, $m_1 = \left\{p+s,p-s,2s,2p\right\}$ and $m = m_0 \cup m_1$.  The local nonlinear force interaction appears as a source term at the second-order perturbation equation and is given by:
\begin{equation}
\begin{array}{l}
\displaystyle \sum_{m_1}\hat{f}_2^{({m_1})}\left(q^{(0,s)}_{X_0},q^{(0,p)}_{X_0}\right) = \hat{f}_{(2,0)}+\hat{f}_{(2,\bar{\omega}_p-\bar{\omega}_s)}+\hat{f}_{(2,2\bar{\omega}_s)}+\hat{f}_{(2,\bar{\omega}_p+\bar{\omega}_s)}\cdots \\
+\hat{f}_{(2,2\bar{\omega}_p)}+\hat{f}_{(2,-\bar{\omega}_p+\bar{\omega}_s)}+\hat{f}_{(2,-2\bar{\omega}_s)}+\hat{f}_{(2,-(\bar{\omega}_p+\bar{\omega}_s))}+\hat{f}_{(2,-2\bar{\omega}_p)},
\end{array}
\end{equation}
where
\begin{equation}
\begin{array}{l}
	\hat{f}_{(2,0)} =  2\left(\Delta \Phi_s^2 |\hat{U}_s|^2e^{2\Im(\mu_{s})X_0} + \Delta \Phi_p^2 |\hat{U}_p|^2\right),\\
	\hat{f}_{(2,2\bar{\omega}_s)} =  \Delta \Phi_s^2 \hat{U}_s^{2} e^{{\tt i} (2\bar{\omega}_{s}T_0-2\mu_s X_0)},\\
	\hat{f}_{(2,2\bar{\omega}_p)} =  \Delta \Phi_p^2 \hat{U}_p^{2} e^{{\tt i} (2\bar{\omega}_{p}T_0-2\mu_p X_0)},\\
	\hat{f}_{(2,\bar{\omega}_p-\bar{\omega}_s)} =  2 \Delta \Phi_s \Delta \Phi_p \hat{U}_s^{\ast}\hat{U}_p e^{{\tt i} ((\bar{\omega}_{p}-\bar{\omega}_s)T_0-(\mu_p-\mu_s^{\ast}) X_0)},\\
	\hat{f}_{(2,\bar{\omega}_p+\bar{\omega}_s)} =  2 \Delta \Phi_s \Delta \Phi_p \hat{U}_s\hat{U}_p e^{{\tt i} ((\bar{\omega}_{p}+\bar{\omega}_s)T_0-(\mu_p+\mu_s) X_0)},\\
	\hat{f}_{(2,-2\bar{\omega}_s)} =  \Delta \Phi_s^2 \hat{U}_s^{\ast 2} e^{-{\tt i} (2\bar{\omega}_{s}T_0-2\mu_s^{\ast} X_0)},\\
	\hat{f}_{(2,-2\bar{\omega}_p)} =  \Delta \Phi_p^2 \hat{U}_p^{\ast 2} e^{-{\tt i} (2\bar{\omega}_{p}T_0-2\mu_p^{\ast} X_0)},\\
	\hat{f}_{(2,-\bar{\omega}_p+\bar{\omega}_s)} =  2 \Delta \Phi_s \Delta \Phi_p \hat{U}_s\hat{U}_p^{\ast} e^{{\tt i} (-(\bar{\omega}_{p}-\bar{\omega}_s)T_0-(\mu_s-\mu_p^{\ast}) X_0)},\\
	\hat{f}_{(2,-\bar{\omega}_p-\bar{\omega}_s)} =  2 \Delta \Phi_s \Delta \Phi_p \hat{U}_s^{\ast}\hat{U}_p^{\ast} e^{{\tt i} (-(\bar{\omega}_{p}+\bar{\omega}_s)T_0+(\mu_p^{\ast}+\mu_s^{\ast}) X_0)},
\end{array}
\end{equation}
with $\Delta\Phi_{m_0} = \boldsymbol{\Phi}_{m_0}(2)-\boldsymbol{\Phi}_{m_0}(1)$, and the superscript $(~^{\ast}~)$ denotes the complex conjugate operator.
  
As the analysis is restricted to the frequencies $\bar{\omega}_s \approx \bar{\omega}_{\tt R}$ and $\bar{\omega}_p \approx 2 \bar{\omega}_{\tt R}$, where 2:1 internal resonance is expected to occur, terms with harmonic $2(\bar{\omega}_sT_0-\mu_s X_0)$ can be resonant with the harmonic $(\bar{\omega}_pT_0-\mu_p X_0)$, while terms with harmonic $((\bar{\omega}_p-\bar{\omega}_s)T_0-(\mu_{p}-\mu^{\ast}_s) X_0)$ are resonant with $(\bar{\omega}_s T_0-\mu_s X_0)$. Using Eqs. \eqref{eq:detuning_p} and \eqref{eq:detuning_s}, these harmonics are related as follows:
\begin{subequations}
	\begin{align}	
	2(\bar{\omega}_sT_0-\mu_s X_0) &=\left(2\bar{\omega}_{\tt R}T_0-\mu_s X_0\right)+2\sigma_s T_1 \\\nonumber
	  &=\left(\bar{\omega}_pT_0-\mu_p X_0\right) -\left(\sigma_p-2\sigma_s\right) T_1 + \sigma_{\mu_r} X_1 + {\tt i} \sigma_{\mu_i} X_1,\\
	((\bar{\omega}_p-\bar{\omega}_s)T_0-(\mu_{p}-\mu^{\ast}_s) X_0) &= \bar{\omega}_{\tt R} T_0+\left(\sigma_p-\sigma_s\right) T_1  -(\mu_{p}-\mu^{\ast}_s) X_0\\\nonumber
	  &=\left(\bar{\omega}_sT_0-\mu_s X_0\right)+\left(\sigma_p-2\sigma_s\right) T_1- \sigma_{\mu_r} X_1,
	\end{align}
\end{subequations}
where $\sigma_{\mu_r}$ and $\sigma_{\mu_i}$ are given by
\begin{subequations}
	\begin{align}
	\sigma_{\mu_r} =& \varepsilon^{-1}\left(\Re(\mu_p)-2\Re(\mu_s)\right),\\
	\sigma_{\mu_i} =& \varepsilon^{-1}\left(\Im(\mu_p)-2\Im(\mu_s)\right),
	\end{align}
\end{subequations}
with $\mu_p$ and $\mu_s$ determined from the dispersion relation given in Eq. \eqref{eq:discrete_linear_e7}.

Therefore, in Eq. \eqref{eq:app_MMS_o2}, the non-homogeneous terms with harmonics $(\bar{\omega}_pT_0-\mu_p X_0)$, $(\bar{\omega}_sT_0-\mu_s X_0)$, $2(\bar{\omega}_sT_0-\mu_s X_0)$ and $((\bar{\omega}_p-\bar{\omega}_s)T_0-(\mu_{p}-\mu^{\ast}_s) X_0)$ contribute to the secular terms and must vanish in order to provide a uniform expansion. Then, the following equations must hold:  
\begin{equation} 
\begin{array}{l}
	\bar{\omega}_p {\bf M}D_{T_1} \hat{\bf u}^{(0,p)}_j-\mu_p\gamma_c^2\displaystyle \sum_{q = -1}^{0}{\bf K}_q^{1} {\hat{\bf u}}^{(0,p)}_{j+q}=-\bar{\omega}_p\hat{\zeta}_m\gamma_c {\displaystyle\sum_{q = -1}^{1}}{\bf C}_q {\hat{\bf u}}^{(0,p)}_{j+q}+{\tt i}\frac{ \hat{\gamma}_2}{2} \boldsymbol{\Phi}_{f} \hat{f}_{(2,2\omega_s)},\\
	\bar{\omega}_s {\bf M}D_{T_1} \hat{\bf u}^{(0,s)}_j-\mu_s\gamma_c^2\displaystyle \sum_{q = -1}^{0}{\bf K}_q^{1} {\hat{\bf u}}^{(0,s)}_{j+q}=-\bar{\omega}_s\hat{\zeta}_m\gamma_c {\displaystyle\sum_{q = -1}^{1}}{\bf C}_q {\hat{\bf u}}^{(0,s)}_{j+q}+{\tt i}\frac{ \hat{\gamma}_2}{2} \boldsymbol{\Phi}_{f} \hat{f}_{(2,\omega_p-\omega_s)},    
\end{array}
	\label{eq:app_MMS_solvability}
\end{equation}
Substituting the expanded form of $\hat{\bf u}_{j}^{(0,p)}$ and $\hat{\bf u}_j^{(0,s)}$ provided in Eq. \eqref{eq:app_u_0order} into \eqref{eq:app_MMS_solvability}, and left multiplying them by the corresponding transposed wave mode shape vector yield the following solvability conditions at order $\varepsilon^2$:
\begin{subequations}
	\begin{align}
	D_{T_1}\hat{U}^{(p)}-\nu_p D_{X_1}\hat{U}^{(p)} =& -\hat{c}_p\hat{U}^{(p)}+{\tt i}\alpha_p \hat{U}^{(s)2} e^{-\sigma_{\mu_i}X_1}e^{-{\tt i}\theta(T_1,X_1)},
	\label{eq:discrete_e25a}\\
	D_{T_1}\hat{U}^{(s)}-\nu_s D_{X_1}\hat{U}^{(s)} =& -\hat{c}_s\hat{U}^{(s)}+{\tt i}\alpha_s \hat{U}^{(p)} \hat{U}^{(s)\ast}e^{+\sigma_{\mu_i}X_1} e^{{\tt i}\theta(T_1,X_1)},
	\label{eq:discrete_e25b}
	\end{align}
	\label{eq:discrete_e25bb}
\end{subequations}
where 
\begin{equation*}
	\alpha_p = \hat{\gamma}_{2p}\boldsymbol{\Phi}_p^T \boldsymbol{\Phi}_f \Delta\boldsymbol{\Phi}_s^2, \quad \alpha_s = \hat{\gamma}_{2s}\boldsymbol{\Phi}_s^T \boldsymbol{\Phi}_f \Delta\boldsymbol{\Phi}_p\Delta\boldsymbol{\Phi}_s,
\end{equation*}
\begin{equation*}
\nu_r = \frac{\mu_r \gamma_c^2}{\bar{\omega}_r m_r},  \:
\hat{\gamma}_{2r} = \frac{\hat{\gamma}_2}{2\bar{\omega}_r m_r}, \: 
m_r = \boldsymbol{\Phi}_r^T{\bf M}\boldsymbol{\Phi}_r, \:
\hat{c}_r = \frac{\hat{\zeta}_m\gamma_c}{m_r}\boldsymbol{\Phi}_r^T\displaystyle \sum_{q=-1}^{1} \hat{\bf C}_q\boldsymbol{\Phi}_r \quad
\text{for } r = \left\{p,s\right\},
\end{equation*}
and
\begin{equation*}
\theta(T_1,X_1) = (\sigma_p-2\sigma_s)T_1-\sigma_{\mu_r} X_1.
\end{equation*}
Assuming complex wave mode amplitudes in the form:
\begin{equation}
\hat{U}^{(p)} = \hat{U}_{R}^{(p)}+{\tt i}\hat{U}_{I}^{(p)}, \quad \hat{U}^{(s)} = \hat{U}_{R}^{(s)}+{\tt i}\hat{U}_{I}^{(s)},
\label{eq:discrete_e33}
\end{equation}
and equating real and imaginary parts in Eqs. \eqref{eq:discrete_e25a} and \eqref{eq:discrete_e25b} yield the following system of differential equations depending on the nature of the interacting wave modes:
\begin{subequations}
	\begin{align}
	\nonumber
	\text{case propagative-propagative:}&&\\	
	\nonumber
	D_{T_1}\hat{U}_R^{(p)}-\nu_p D_{X_1}\hat{U}_R^{(p)}=& -\hat{c}_p\hat{U}_R^{(p)} +{\alpha}_p \left\{\left(\hat{U}_R^{(s)2}-\hat{U}_I^{(s)2}\right)\sin \theta+\cdots\right. \\
	&\left.-2\hat{U}_R^{(s)}\hat{U}_I^{(s)} \cos \theta\right\},\\
	\nonumber
	D_{T_1}\hat{U}_I^{(p)}-\nu_p D_{X_1}\hat{U}_I^{(p)}=& -\hat{c}_p\hat{U}_I^{(p)} +{\alpha}_p \left\{\left(\hat{U}_R^{(s)2}-\hat{U}_I^{(s)2}\right)\cos \theta+\cdots\right. \\
	&\left.+2\hat{U}_R^{(s)}\hat{U}_I^{(s)} \sin \theta\right\},\\
	\nonumber
	D_{T_1}\hat{U}_R^{(s)}-\nu_s D_{X_1}\hat{U}_R^{(s)}=& -\hat{c}_s\hat{U}_R^{(s)} -{\alpha}_s \left\{\left(\hat{U}_I^{(p)}\hat{U}_R^{(s)}-\hat{U}_R^{(p)}\hat{U}_I^{(s)}\right)\cos \theta+\cdots\right.\\
	&\left.+\left(\hat{U}_R^{(p)}\hat{U}_R^{(s)}+\hat{U}_I^{(p)}\hat{U}_I^{(s)}\right)\sin \theta\right\},\\
	\nonumber
	D_{T_1}\hat{U}_I^{(s)}-\nu_s D_{X_1}\hat{U}_I^{(s)}=& -\hat{c}_s\hat{U}_I^{(s)} +{\alpha}_s \left\{\left(\hat{U}_R^{(p)}\hat{U}_R^{(s)}+\hat{U}_I^{(p)}\hat{U}_I^{(s)}\right)\cos \theta+\cdots\right.\\ 
	&\left.+\left(\hat{U}_I^{(p)}\hat{U}_R^{(s)}-\hat{U}_R^{(p)}\hat{U}_I^{(s)}\right)\sin \theta\right\},\\
	\nonumber
	\text{case propagative-evanescent:}&&\\	
	\nonumber
	D_{T_1}\hat{U}_R^{(p)}-\nu_p D_{X_1}\hat{U}_R^{(p)}=& -\hat{c}_p\hat{U}_R^{(p)} +{\alpha}_p e^{-\sigma_{\mu_i}X_1}\left\{\left(\hat{U}_R^{(s)2}-\hat{U}_I^{(s)2}\right)\sin \theta+\cdots\right. \\
	&\left.-2\hat{U}_R^{(s)}\hat{U}_I^{(s)} \cos \theta\right\},\\
	\nonumber
	D_{T_1}\hat{U}_I^{(p)}-\nu_p D_{X_1}\hat{U}_I^{(p)}=& -\hat{c}_p\hat{U}_I^{(p)} +{\alpha}_p e^{-\sigma_{\mu_i}X_1}\left\{\left(\hat{U}_R^{(s)2}-\hat{U}_I^{(s)2}\right)\cos \theta+\cdots\right. \\
	&\left.+2\hat{U}_R^{(s)}\hat{U}_I^{(s)} \sin \theta\right\},\\
	\nonumber
	D_{T_1}\hat{U}_R^{(s)}+\sgn(\nu_s)|\nu_s| D_{X_1}\hat{U}_I^{(s)}=& -\hat{c}_s\hat{U}_R^{(s)} -{\alpha}_s \left\{\left(\hat{U}_I^{(p)}\hat{U}_R^{(s)}-\hat{U}_R^{(p)}\hat{U}_I^{(s)}\right)\cos \theta+\cdots\right.\\
	&\left.+\left(\hat{U}_R^{(p)}\hat{U}_R^{(s)}+\hat{U}_I^{(p)}\hat{U}_I^{(s)}\right)\sin \theta\right\},\\
	\nonumber
	D_{T_1}\hat{U}_I^{(s)}-\sgn(\nu_s)|\nu_s| D_{X_1}\hat{U}_R^{(s)}=& -\hat{c}_s\hat{U}_I^{(s)} +{\alpha}_s \left\{\left(\hat{U}_R^{(p)}\hat{U}_R^{(s)}+\hat{U}_I^{(p)}\hat{U}_I^{(s)}\right)\cos \theta+\cdots\right.\\ 
	&\left.+\left(\hat{U}_I^{(p)}\hat{U}_R^{(s)}-\hat{U}_R^{(p)}\hat{U}_I^{(s)}\right)\sin \theta\right\}.
	\end{align}
	\label{eq:MMS_ODE_system}
\end{subequations}

Indeed, Eq. \eqref{eq:MMS_ODE_system} can be written in matrix form as follows:
\begin{equation}
	D_{T_1}\hat{\bf a} - {\bf V}D_{X_1}\hat{\bf a} = -{\bf Q}\hat{\bf a}+\hat{\bf q}_{\tt NL},
	\label{eq:advection_eq} 
\end{equation}
where 
\begin{equation*}
	\begin{array}{l}
{\bf V} = \begin{bmatrix}
		\nu_p & 0 & 0 & 0\\
		0 & \nu_p & 0 & 0\\
		0 & 0 & \nu_s & 0\\
		0 & 0 & 0 & \nu_s 
	\end{bmatrix}, \quad \text{in the propagative-propagative case, or}\\
~\\	
{\bf V} = \begin{bmatrix}
		\nu_p & 0 & 0 & 0\\
		0 & \nu_p & 0 & 0\\
		0 & 0 & 0 & -\sgn(\nu_s)|\nu_s|\\
		0 & 0 & \sgn(\nu_s)|\nu_s| & 0 
	\end{bmatrix}, \quad \text{in the propagative-evanescent case,}
\end{array}
\end{equation*}
and
\begin{equation*}
	\begin{array}{l}
	\hat{\bf a} = 
	\begin{bmatrix}
		\hat{U}^{(p)}_R\\
		\hat{U}^{(p)}_I\\
		\hat{U}^{(s)}_R\\
		\hat{U}^{(s)}_I
	\end{bmatrix}, \quad
	{\bf Q} = \begin{bmatrix}
		\hat{c}_p & 0 & 0 & 0\\
		0 & \hat{c}_p & 0 & 0\\
		0 & 0 & \hat{c}_s & 0\\
		0 & 0 & 0 & \hat{c}_s
	\end{bmatrix},\\ ~\\
	\hat{\bf q}_{\tt NL} = 
	\begin{bmatrix}
		+{\alpha}_p e^{-\sigma_{\mu_i}X_1}\left\{\left(\hat{U}_R^{(s)2}-\hat{U}_I^{(s)2}\right)\sin \theta-2\hat{U}_R^{(s)}\hat{U}_I^{(s)} \cos \theta\right\},\\
		+{\alpha}_p e^{-\sigma_{\mu_i}X_1}\left\{\left(\hat{U}_R^{(s)2}-\hat{U}_I^{(s)2}\right)\cos \theta+2\hat{U}_R^{(s)}\hat{U}_I^{(s)} \sin \theta\right\}\\
		-{\alpha}_s \left\{\left(\hat{U}_I^{(p)}\hat{U}_R^{(s)}-\hat{U}_R^{(p)}\hat{U}_I^{(s)}\right)\cos \theta+\left(\hat{U}_R^{(p)}\hat{U}_R^{(s)}+\hat{U}_I^{(p)}\hat{U}_I^{(s)}\right)\sin \theta\right\}\\
		+{\alpha}_s \left\{\left(\hat{U}_R^{(p)}\hat{U}_R^{(s)}+\hat{U}_I^{(p)}\hat{U}_I^{(s)}\right)\cos \theta+\left(\hat{U}_I^{(p)}\hat{U}_R^{(s)}-\hat{U}_R^{(p)}\hat{U}_I^{(s)}\right)\sin \theta\right\}
	\end{bmatrix}.
	\end{array}
\end{equation*}
In the case of propagative-propagative interaction, these differential equations in \eqref{eq:advection_eq} are correspond to 1D advection equations with nonlinear source term. Instead, they are ``quasi''-advection equations with nonlinear source term in the case of propagative-evanescent wave-wave interaction. The term ``quasi'' comes from the fact that due to the evanescent nature of the half subharmonic wave mode into consideration, $\nu_s$ is imaginary and the advection coefficient matrix ${\bf V}$ is not diagonal, it contains off-diagonal terms, which couples the imaginary and real parts of the subharmonic components of the advection term. Physically, this is due to the fact that the evanescent wave does not propagate, thus the energy from the active part (real component) can only be transferred to the reactive (imaginary component) and vice-versa in space.     

\subsection{Free wave propagation}
Using the multiple scales framework derived above, a solution for a free wave propagation problem involving wave-wave interaction is sought in this section. In this case, initial conditions should be considered, as for instance:
\begin{equation}
	{\bar{u}}_j(0) = u_e(j),
	\label{eq:MMS_free_e1}
\end{equation}
where $u_e(\tau)$ is a general shape function. 

Within the framework of the method of multiple scales, the problem to be solved at the different orders of perturbation in the present case is given by:
\begin{subequations}
	\begin{align}
	\varepsilon^1:\quad&
	{\bf M} D_{T_0}^2 \hat{\bf u}^{(0)}_{j}+ \gamma_c^2 {\displaystyle \sum_{p = -1}^{1}}{\bf K}_p {\hat{\bf u}}^{(0)}_{j+p}= {\bf 0}, &
	\label{eq:MMS_free_e2a}\\
	\varepsilon^2:\quad&
    {\bf M}D_{T_0}^2 \hat{\bf u}^{(1)}_{j}+ \gamma_c^2 {\displaystyle \sum_{p = -1}^{1}}{\bf K}_p {\hat{\bf u}}^{(1)}_{j+p} = -2{\bf M}D_{T_1}D_{T_0} \hat{\bf u}^{(0)}_{j}-2\hat{\zeta}_m\gamma_c {\displaystyle\sum_{p = -1}^{1}}{\bf C}_p D_{T_0}{\hat{\bf u}}^{(0)}_{X_0+p}&\\ \nonumber
    &-2\gamma_c^2D_{X_0}\left(e^{-{\tt i}\mu_0 X_0}\right){\displaystyle \sum_{p = -1}^{0}{\bf K}_p^{1}{\tilde{\bf u}}^{(0)}_{j+p}}- \hat{\gamma}_2 \boldsymbol{\Phi}_{f} \left(\hat{u}^{(0)}_{j}-\hat{v}^{(0)}_{j}\right)^2,&
    \label{eq:MMS_free_e2b}\\
	\nonumber
	\text{subject to:}&\\
	&{\hat{u}}_{j}^{(0)}(0) = \varepsilon^{-1}u_e(j). &
	\end{align}	
	\label{eq:MMS_free_e2}
\end{subequations}

Considering the initial shape (initial condition) can be decomposed in two wave components, as follows:
\begin{equation}
 	u_e(X_0) = U_e^{(p)} e^{-{\tt i}\mu_p X_0+\phi_p}+ U_e^{(s)} e^{-{\tt i}\mu_sX_0+\phi_s},
\end{equation}
where $\bar{\omega}_p$ is the normalized angular frequency of excitation (or primary frequency), $U_e^{(p)}$, $\phi_p$ and $U_e^{(s)}$, $\phi_s$ are the magnitudes and phases of the wave shapes relative the primary and subharmonic components, respectively, with $U_e^{(s)} \ll U_e^{(p)}$, and c.c. stands for the complex conjugate. 

Then, the zeroth-order solution to the problem should satisfy the solvability conditions under the constraints imposed by the initial condition, which applied to the system of evolution equations (Eq. \eqref{eq:MMS_ODE_system}) become:
\begin{subequations}
	\begin{align}
		\hat{U}^{(p)}_R(X_1,0) = \varepsilon^{-1}U_e^{(p)}\cos\phi_e,& \quad & \hat{U}^{(p)}_I(X_1,0) = \varepsilon^{-1}U_e^{(p)}\sin\phi_e,\\
		\hat{U}^{(s)}_R(X_1,0) = \varepsilon^{-1}U_e^{(s)}\cos\phi_e,& \quad & \hat{U}^{(s)}_I(X_1,0) = \varepsilon^{-1}U_e^{(s)}\sin\phi_e.
	\end{align}
\end{subequations}
   
\subsection{Forced wave propagation}
In this section, a solution by the method of multiple scales is sought for a forced wave propagation problem. In the case of forced wave propagation, the problem involves not only a set of initial conditions, but also boundary conditions. In this paper, the following conditions are considered:
\begin{subequations}
	\begin{align}
	&{\bar{u}}_0(\tau) = u_e(\tau),\\
	&{\bar{u}}_j(0) = 0,\: \dot{\bar{u}}_j(0) = 0, 
	\end{align}
	\label{eq:MMS_forced_e1}
\end{subequations}
where $u_e(\tau)$ is a general temporal excitation. 

Within the framework of the method of multiple scales, the problem to be solved at the different orders of perturbation in the present case is given by:
\begin{subequations}
	\begin{align}
	\varepsilon^1:\quad&
	{\bf M} D_{T_0}^2 \hat{\bf u}^{(0)}_j+ \gamma_c^2 {\displaystyle \sum_{p = -1}^{1}}{\bf K}_p {\hat{\bf u}}^{(0)}_{j+p}= {\bf 0}, &
	\label{eq:MMS_forced_e7a}\\
	\varepsilon^2:\quad&
    {\bf M}D_{T_0}^2 \hat{\bf u}^{(1)}_j+ \gamma_c^2 {\displaystyle \sum_{p = -1}^{1}}{\bf K}_p {\hat{\bf u}}^{(1)}_{j+p} = -2{\bf M}D_{T_1}D_{T_0} \hat{\bf u}^{(0)}_{j}-2\hat{\zeta}_m\gamma_c {\displaystyle\sum_{p = -1}^{1}}{\bf C}_p D_{T_0}{\hat{\bf u}}^{(0)}_{j+p}&\\ \nonumber
    &-2\gamma_c^2D_{X_0}\left(e^{-{\tt i}\mu_0 X_0}\right){\displaystyle \sum_{p = -1}^{0}{\bf K}_p^{1}{\tilde{\bf u}}^{(0)}_{j+p}}- \hat{\gamma}_2 \boldsymbol{\Phi}_{f} \left(\hat{u}^{(0)}_{j}-\hat{v}^{(0)}_{j}\right)^2,&
    \label{eq:MMS_forced_e7c}\\
	\nonumber
	\text{subject to:}&\\
	&{\hat{\bf u}}_0^{(0)}(\tau) = \varepsilon^{-1}{\bf u}_e(\tau),&\\
	&{\hat{\bf u}}_j^{(0)}(0) = {\bf 0},\: \dot{\hat{\bf u}}_j^{(0)}(0) = {\bf 0}. &
	\end{align}	
	\label{eq:MMS_forced_e7}
\end{subequations}\linebreak

Let's consider the localized excitation (boundary condition) is given by a time-harmonic displacement applied at $j = 0$, {\it i.e.}
\begin{equation}
 	{\bf u}_e(\tau) = U_e \boldsymbol{\Phi}_p e^{{\tt i}\bar{\omega}_p\tau+\phi_p}+\varepsilon U_e \boldsymbol{\Phi}_s e^{{\tt i}\bar{\omega}_s\tau+\phi_s},
\end{equation}
where $\bar{\omega}_p$ is the normalized angular frequency of excitation (or primary frequency), $U_e^{(p)}$, $\phi_p$ and $U_e^{(s)}$, $\phi_s$ are the magnitude and phase angles relative to the primary and subharmonic components, respectively, with $U_e^{(s)} \ll U_e^{(p)}$, and c.c. stands for the complex conjugate. 

In the case of a forced problem, the zeroth-order solution to the problem should satisfy the solvability conditions under the constraints imposed by the initial and boundary conditions (Eq. \eqref{eq:MMS_forced_e7}), which applied to Eq. \eqref{eq:MMS_ODE_system} become:
\begin{subequations}
	\begin{align}
		\hat{U}^{(p)}_R(0,T_1) &= \varepsilon^{-1} U_e^{(p)} \cos \phi_p, & \quad & \hat{U}^{(p)}_R(X_1,0) = 0,\\
		\hat{U}^{(p)}_I(0,T_1) &=  \varepsilon^{-1} U_e^{(p)} \sin \phi_p,& \quad & \hat{U}^{(p)}_I(X_1,0) = 0,\\
		\hat{U}^{(s)}_R(0,T_1) &= \varepsilon^{-1} U_e^{(s)} \cos \phi_s, & \quad & \hat{U}^{(s)}_R(X_1,0) = 0,\\
		\hat{U}^{(s)}_I(0,T_1) &= \varepsilon^{-1} U_e^{(s)} \sin \phi_s, & \quad & \hat{U}^{(s)}_I(X_1,0) = 0.
	\end{align}
\end{subequations}

\subsection{Numerical implementation}
To numerically solve the set of first-order reaction-advection equations describing the dynamics of the slow-long wave modulations of a locally resonant metamaterial with quadratic local interaction, one can make use of a pseudo-discretization, the so-called method of lines. It consists in discretizing only the spatial component of the system of partial differential equations, transforming the problem to a set of ordinary differential equations (ODEs) to be solved using a standard ODE integration technique. Herein, our interest is in the dynamics of right-going waves, in which $\nu_p > 0$ is positive and $\nu_s$ is either positive (case propagative-propagative) or purely imaginary, which means zero velocity. Thus, using finite differences, a downwind spatial discretization is employed. For the temporal discretization, the implicit backward Euler scheme is used in association with a Newton-Raphson procedure, to solve the residue equation including nonlinear source terms. %A scheme of the numerical implementation is presented in Fig. \ref{fig:}.     

%\section{Results} 
%\subsection{Free wave propagation}

%\subsection{Forced wave propagation}

%\subsubsection{Propagative-Propagative wave-wave interaction}

%\subsubsection{Propagative-Evanescent wave-wave interaction}

%\subsection{Effect of the subwavelength nature on the wave-wave interaction}

\section{Conclusions}
In this paper, it was shown that the consideration of nonlinearity in resonant inclusions of metamaterials bring about interaction between propagating and evanescent waves. Using a perturbation method in terms of multiple time and space scales, the possibility of energy exchange between these waves is revealed. This nonlinear mechanism is responsible for emergent attenuation zones in the metamaterial that do not occur in the linear regime.   

%\bibliography{Paper4_references}

\appendix
\section{Consequences of considering multiple space scales}
\label{app:A}

Let's assume a linear and uniform dispersive medium. The relation between the non-dimensional wavenumber $\mu_0$ and the angular frequency $\bar{\omega}_0$ in this case is given by:
\begin{equation}
	\mu_0 = \mu_0(\bar{\omega}_0),
\end{equation}
where $\omega_0 \in \Re$ and $\mu_0$ might be complex.

Assuming a perturbation in both the wavenumber and the angular frequency. They can be written in series expansion as:
\begin{subequations}
	\begin{align}
		\mu &= \mu_0 + \varepsilon \mu_1 + \varepsilon^2 \mu_2 + \cdots,\\
		\bar{\omega} &= \bar{\omega}_0 + \varepsilon\bar{\omega}_1 + \varepsilon^2 \bar{\omega}_2 + \cdots
	\end{align}
\end{subequations}

Considering two time and space scales within the framework of the method of multiple scales, such that $T_0 = \tau$, $T_1 = \varepsilon\tau$, $X_0 = j$, $X_1 = \varepsilon j$, with $\varepsilon \ll 1$ being a small dimensionless parameter, the zeroth-order solution for a problem can be written as:
\begin{equation}
	u_j^{(0)} = U_0(X_1,T_1) e^{{\tt i}(\bar{\omega}_0 T_0-\mu_0 X_0)},
	\label{eq:u0}
\end{equation}
where $U_0(X_1,T_1)$ is the complex wave modulation function. 

Assuming the long space and slow time are independent variables, one can express the complex wave modulation as:
\begin{equation}
	U_0(X_1,T_1) = \bar{U}_0 e^{{\tt i}\bar{\omega}_1 T_1} e^{-{\tt i} \mu_1 X_1}
	\label{eq:U1}
\end{equation}

The zeroth-order approximation of the neighbouring interaction among chain masses is described by:
\begin{equation}
	f_{x}^{(0)} = - u^{(0)}_{j-1}+2u^{(0)}_j - u^{(0)}_{j+1}.
	\label{eq:f_x}
\end{equation}
Substituting the zeroth-order solution (Eqs. \eqref{eq:u0} and \eqref{eq:U1}) in \eqref{eq:f_x}, yields:
\begin{equation}
	f_{x}^{(0)} = \bar{U}_0e^{{\tt i}(\bar{\omega}_0 T_0+\bar{\omega}_1 T_1)}\left(-e^{{\tt i} \mu_1}e^{{\tt i} \mu_0}+2-e^{-{\tt i} \mu_1}e^{-{\tt i} \mu_0}\right).
\end{equation}
Taylor expansion of a exponential function about zero provides:
\begin{equation}
	e^{\varepsilon x} = 1 + \varepsilon x + \mathcal{O}(\varepsilon^2 x^2).
	\label{eq:exp_approx}
\end{equation}
Assuming $\mu_1$ small, a good approximation for $f_{x}^{(0)}$ is given by:
\begin{equation}
	\begin{array}{ll}
	f_{x}^{(0)} &\simeq \bar{U}_0e^{{\tt i}(\bar{\omega}_0 T_0+\bar{\omega}_1 T_1)}\left(-(1+{\tt i}\varepsilon\bar{\mu}_1)e^{{\tt i}\mu_0}+2-(1-{\tt i}\varepsilon\bar{\mu}_1)e^{-{\tt i}\mu_0}\right)\\
	~ &\simeq \bar{U}_0e^{{\tt i}(\bar{\omega}_0 T_0+\bar{\omega}_1 T_1)}\left((2-e^{{\tt i}\mu_0}-e^{-{\tt i}\mu_0})+{\tt i}\varepsilon\bar{\mu}_1(e^{-{\tt i}\mu_0}-e^{{\tt i}\mu_0})\right).
	\end{array}
	\label{eq:fdisp_1}
\end{equation}
Using a central difference scheme, the following are valid:
\begin{subequations}
	\begin{align}
		D_{X_0}\left(e^{-{\tt i}\mu_0 X_0}\right) =&  \frac{1}{2}(e^{-{\tt i}\mu_0}-e^{{\tt i}\mu_0}),\\
		D^2_{X_0}\left(e^{-{\tt i}\mu_0 X_0}\right) =& (+e^{{\tt i}\mu_0}-2+e^{-{\tt i}\mu_0}).
	\end{align}
	\label{eq:diff_fin}
\end{subequations}
Then, substituting \eqref{eq:diff_fin} into \eqref{eq:fdisp_1}, yields:
\begin{equation}
	f_{x}^{(0)} \simeq \bar{U}_0e^{{\tt i}(\bar{\omega}_0 T_0+\bar{\omega}_1 T_1)}\left(-D^2_{X_0}\left(e^{-{\tt i}\mu_0 X_0}\right)-2(-{\tt i}\mu_1\varepsilon)D_{X_0}\left(e^{-{\tt i}\mu_0 X_0}\right)\right).
\end{equation}
Using \eqref{eq:exp_approx} and \eqref{eq:diff_fin}, one can write:
\begin{equation}
	(-{\tt i}\mu_1\varepsilon) \simeq \frac{1}{2}\left(e^{-{\tt i}\mu_1 \varepsilon}-e^{+{\tt i}\mu_1 \varepsilon}\right) = \varepsilon D_{X_1}\left(e^{-{\tt i}\mu_1 X_1}\right).
\end{equation}
Thus, the neighbouring interaction force can be expressed as:
\begin{equation}
	f_{x}^{(0)} \simeq \bar{U}_0e^{{\tt i}(\bar{\omega}_0 T_0+\bar{\omega}_1 T_1)}\left(-D^2_{X_0}\left(e^{-{\tt i}\mu_0 X_0}\right)-\varepsilon 2D_{X_0}\left(e^{-{\tt i}\mu_0 X_0}\right)D_{X_1}\left(e^{-{\tt i}\mu_1 X_1}\right)\right).
\end{equation}

%\section{Response in terms of wave modes by a localized excitation}
%\label{app:B}

\end{document}